\documentclass[conference]{IEEEtran}

\usepackage{cite}
\usepackage{amsmath,amssymb,amsfonts}
\usepackage[ruled,linesnumbered,algo2e]{algorithm2e}
\usepackage{graphicx}
\usepackage{textcomp}
\usepackage{tikz}
\usetikzlibrary{arrows.meta, positioning, shapes.geometric, fit, backgrounds}
\usepackage{xcolor}
\usepackage{url}
\usepackage{verbatim}

\usepackage{booktabs}
\usepackage{multirow}
\usepackage{colortbl}
\usepackage{listings}
\usepackage{xcolor}

\definecolor{keywordcolor}{rgb}{0.0, 0.2, 0.6}
\definecolor{commentcolor}{rgb}{0.0, 0.5, 0.0}
\definecolor{stringcolor}{rgb}{0.6, 0.1, 0.1}
\definecolor{outbgcolor}{rgb}{0.95, 0.95, 0.95}

\lstdefinestyle{pythonstyle}{
    language=Python,
    basicstyle=\ttfamily\footnotesize,
    keywordstyle=\bfseries\color{keywordcolor},
    commentstyle=\itshape\color{commentcolor},
    stringstyle=\color{stringcolor},
    showstringspaces=false,
    frame=tb,
    numbers=left,
    numberstyle=\tiny\color{gray},
    numbersep=5pt,
    breaklines=true,
    captionpos=b,
    xleftmargin=1.2em,
    abovecaptionskip=5pt,
    belowcaptionskip=5pt
}

\lstdefinestyle{outputstyle}{
    basicstyle=\ttfamily\footnotesize,
    backgroundcolor=\color{outbgcolor},
    frame=single,
    rulecolor=\color{gray!40},
    breaklines=true,
    xleftmargin=0em,
    numbers=none
}

\usepackage{zi4}

\newtheorem{definition}{Definition}

\begin{document}

\title{Scalable Maximal Frequent Episode Mining with Desbordante}
\date{}

\author{
\IEEEauthorblockN{Maxim Ivanov\textsuperscript{1}, Matvei Smirnov\textsuperscript{1}, Alisa Strazdina\textsuperscript{1}, George Chernishev\textsuperscript{1, 2}}
\IEEEauthorblockA{
\textsuperscript{1}Desbordante Team \\ Saint Petersburg, Russia \\
\textsuperscript{2}Saint Petersburg State University \\
Saint Petersburg, Russia\\  \{maxxim.s.ivanovv, matvei.mik.smirnov, alisastrazdina, chernishev\}@gmail.com\\}}
\maketitle

\begin{abstract}

Episode mining aims to extract subsequences of events that possess certain distinctive properties and constitute facts valuable to the user. Maximal frequent episode mining concentrates on discovery of frequently-appearing subsequences, which are not included into any other larger frequent subsequence. The state-of-the-art for this problem is the MaxFEM algorithm which enumerates possible subsequences, while applying various pruning techniques to accelerate the search.

However, this is a computationally-intensive problem: reducing the minimum number of required subsequence occurrences or increasing the length of the subsequence both substantially raise running time, which limits practical use of MaxFEM.

In this paper we describe our efforts in designing a high-performing algorithm for this problem. For this we: 1) develop an efficient C++ implementation of MaxFEM, and 2) devise an efficient technique to parallelizing it. As the result, we propose an improved parallel MaxFEM variant, which we call ParMaxFEM. Additionally, we integrate the improved algorithm into Desbordante~--- a high-performance, open-source data profiler with deep Python integration that treats patterns as first-class entities and allows users to develop their custom programs that can include discovery and validation of patterns.

To evaluate our approach we compare both C++ implementations with the original SPMF implementation. Experiments demonstrated that our reimplemented version provides up to $8\times$ speedup over the SPMF baseline, while our parallelization technique provides up to $35\times$ improvement overall (on 8 cores).
\end{abstract}

\section{Introduction}

Data profiling~\cite{10.5555/3312004} is concerned with extracting various types of information from data. Such information can range from simple statistics (e.g., the number of NULL values in a table column) to complex facts indicating the presence of sophisticated patterns in the data. Extracted facts can be interesting~\cite{10.1145/3703323.3703725} on their own and can also be applied in areas such as data cleaning, data deduplication, anomaly detection, and many others.

Data profiling can be performed on a variety of data types: tabular, transactional, graphs, and sequential. In this paper we focus on the latter.

Sequential data is represented as one or several ordered list of events, where each one is associated with a time of occurrence.

Episode mining~\cite{https://doi.org/10.1002/widm.1524} aims to extract patterns from sequential data. It has a large number of applications such as~\cite{https://doi.org/10.1002/widm.1524}: analysis and anomaly detection in network data, analysis of financial data, web log analysis, and many others.

Episode mining is a well-established field, with the first episode-mining algorithms introduced~\cite{10.1023/A:1009748302351, 10.5555/3001460.3001490} about 30 years ago. Its core task is the frequent episode mining (FEM), which is as follows. Given a single sequence of events, finding the subsequences whose number of occurrences in the input sequence is no less than a specified threshold called the minimum support ($minsup$). Those subsequences are called frequent episodes.

However, solving this problem produces~\cite{10.1007/978-3-031-20992-5-8} a lot of frequent episodes some of which are included in each other.

This leads to the following drawbacks:
\begin{enumerate}
    \item It is inconvenient for a user to analyze a large number of similar episodes~--- from the end-user point of view it is better to work with a single representative subsequence from which the other subsequences can be derived.
    \item These episodes require additional memory and computational effort to store and manipulate them. Experimental results~\cite{10.1007/978-3-031-20992-5-8} show that removing such episodes reduces the overall number of returned episodes $2.5 - 4\times$.
\end{enumerate}

This motivation gives rise to the problem of maximal frequent episode mining: find such frequent episodes, which are not included in any other.

The straightforward approach to this is to employ existing FEM algorithms and filter out non-maximal ones. However, it was shown~\cite{10.1007/978-3-031-20992-5-8} that a specialized algorithm is 10--40\% faster, even without taking into account post-processing phase.

MaxFEM (Maximal Frequent Episode Miner)~\cite{10.1007/978-3-031-20992-5-8} is the state-of-the-art algorithm for mining maximal frequent episodes. It incrementally constructs all subsequences fitting the user-specified requirements on minimum number of subsequence occurrences and maximal length of the subsequence. Several pruning techniques for accelerating the search were proposed. It was implemented in Java and included into the SPMF~\cite{10.1007/978-3-319-46131-1-8} toolkit.

However, mining maximal frequent episodes is a computationally-intensive problem: reducing the minimum number of required subsequence occurrences or increasing the length of the sought subsequence both substantially raise running time, which limits practical use of MaxFEM.

Existing MaxFEM algorithm and its implementation suffers from the following drawbacks:

\begin{enumerate}
    \item \textbf{Java implementation}. While Java is suitable for rapid prototyping, it is harder to fully exploit software and hardware performance with it than with lower-level languages such as C++.
    Our research group has demonstrated many times~\cite{10749955,10143047,10516381,9435469} that merely reimplementing costly pattern discovery algorithm in C++ can provide up to $3\times$ speed boost, while simultaneously reducing memory consumption up to $2\times$.
    \item \textbf{Lack of parallelization}. The original MaxFEM algorithm and its implementation is a single-threaded one. Such a computationally-intensive problem requires making full use of all available resources.
    \item \textbf{Lack of proper Python integration}. The SPMF toolkit can be employed from within Python applications, however, authors provide no built-in interface. Instead, they recommend either invoking SPMF as a subprocess and parsing its output, or using third‑party wrappers that perform the same work under the hood. This method introduces several critical drawbacks such as dependency management, engineering overhead, I/O and memory bottlenecks.
\end{enumerate}

Our goal is to design a high-performing algorithm for the problem of maximal frequent episode mining, which addresses all three points.

We start with reimplementing MaxFEM in C++ to produce the efficient baseline version. For this we employ a number of techniques, such as pre-allocation, move semantics, shared storage. Then we design a parallel version of MaxFEM algorithm, the ParMaxFEM. To this end we employ task-based parallelism using thread pools with dynamic load balancing. Finally, in this paper we also describe our efforts of integrating ParMaxFEM into Desbordante. Desbordante (Spanish for \textit{boundless}) is an \textit{open-source} \textit{high-performance} data profiling tool designed for discovery and validation of complex patterns in data.

Desbordante overcomes the abovementioned limitations by exposing a native Python interface. As the result users can efficiently and conveniently develop their custom programs that include discovery and validation of patterns.

Experiments demonstrated that reimplemented version of MaxFEM provides up to $8\times$ speedup over the SPMF baseline while consuming up to $14\times$ less memory on average. At the same time, ParMaxFEM can improve performance further, up to $35\times$, while sacrificing additional memory for thread-local storage. Furthermore, experiments demonstrated that ParMaxFEM is scalable, benefiting from up to 8 cores.

Overall, the contribution of this paper is the following:

\begin{enumerate}
    \item A design of ParMaxFEM~--- a novel scalable parallel algorithm for the problem of maximal frequent episode mining.

    \item Open-source implementations of both ParMaxFEM and MaxFEM algorithms. It includes an integration of ParMaxFEM algorithm into Desbordante~--- a high-performance data profiler.

    \item A comprehensive experimental study, involving 12 datasets and a large number of parameter combinations.
\end{enumerate}

This paper is organized as follows. In Section~\ref{sec:background} we provide the core definitions, necessary for understanding the content of the paper. The Section~\ref{sec:relwork} presents the context of the study and describes the related work. The original MaxFEM algorithm is described in Section~\ref{sec:maxfem}, while ParMaxFEM is discussed in Section~\ref{sec:improvements}. The evaluation is presented in Section~\ref{sec:eval}. We conclude the paper in Section~\ref{sec:conclusion}.

\section{Background}\label{sec:background}

To prepare for the following discussion, we must formally define the core concepts. These definitions are provided as in the original MaxFEM paper~\cite{10.1007/978-3-031-20992-5-8}.

Let $E = \{i_1, i_2, \ldots , i_m\}$ be a finite set of \underline{events} (also called items or symbols). A subset $X \subseteq E$ is called an \underline{event set}.

A discrete sequence, also called a \underline{complex event sequence}, is defined as a finite ordered list $S = \langle(SE_{t_1}, t_1), (SE_{t_2}, t_2), \ldots , (SE_{t_n}, t_n)\rangle$ of pairs of the form $(SE_{t_i}, t_i)$ where $SE_{t_i} \subseteq E$ is an event set and $t_i$ is an integer representing a timestamp. A sequence is ordered by time, that is for any integers $1 \le i < j \le n$, the relationship $t_i < t_j$ holds.

An event set $SE_{t_i}$ of a sequence contains events that are assumed to have occurred at the same time, and for this reason it is called a \underline{simultaneous event set}.

In the case, where a complex event sequence contains event sets each having only one event, it is said to be a \underline{simple event sequence}.

It is to be noted that the same event can appear multiple times in a sequence (in different event sets). Besides, although the definition of sequence includes timestamps, it can also be used to model sequences that do not have timestamps such as sequence of words by assigning contiguous integers as timestamps (e.g. 1, 2, 3, 4, 5).

A \underline{composite episode} $\alpha$ is an ordered list of \underline{simultaneous event sets}. A composite episode $\alpha$ having $p$ event sets is represented as $\alpha = \langle X_1, X_2, \ldots , X_p\rangle$, where
$X_i \subseteq E$, and $X_i$ is said to appear before $X_j$ for any integers $1 \le i < j \le p$. The \underline{size} of $\alpha$ is defined as $size(\alpha) = \bigcup_{i \in [1,p]} |X_i|$.

A \underline{parallel episode} is a composite episode that contains a single event set.

A \underline{serial episode} is a composite episode where each event set contains one event.

\begin{definition}[Occurrence]
    An \underline{occurrence} of an episode $\alpha = \langle X_1, X_2, \ldots, X_p\rangle$ in a complex event sequence $S = \langle (SE_{t_1}, t_1), (SE_{t_2}, t_2), \ldots , (SE_{t_n}, t_n)\rangle$ is a time interval $[t_s, t_e]$ that satisfies $X_1 \subseteq SE_{z_1}, X_2 \subseteq SE_{z_2}, \ldots , X_p \subseteq SE_{z_w}$ for some integers $t_s = z_1 < z_2 < \ldots < z_w = t_e$. In an occurrence $[t_s, t_e]$, $t_s$ is said to be the start point, while $t_e$ is the end point. The length of an occurrence $[t_s, t_e]$ is defined as $t_e - t_s$. The notation $occSet(\alpha)$ refers to the set of all occurrences of $\alpha$ that have a length that is smaller than some maximum length $winlen$ set by the user.
\end{definition}

\begin{definition}[Head support]
    Let there be a composite sequence $S$ and an episode $\alpha$. The \underline{support} of $\alpha$ in $S$ is defined as $sup(\alpha) = |\{t_s|[t_s, t_e] \in occSet(\alpha)\}|$, that is the number of distinct start points in the occurrence set of $\alpha$~\cite{10.1016/j.is.2007.07.003}.
\end{definition}

\begin{definition}[Mining frequent episodes]
    Given, a complex event sequence $S$, a user-defined threshold $minsup > 0$ and a user-specified window length $winlen > 0$, the \underline{problem} of mining frequent episodes is to enumerate all frequent episodes appearing in $S$. An episode $\alpha$ is \underline{frequent} if $sup(\alpha) \ge minsup$~\cite{10.1016/j.is.2007.07.003}.
\end{definition}

\begin{definition}[Mining maximal frequent episodes in a complex event sequence]
    Given, a complex event sequence $S$, a user-defined threshold $minsup > 0$ and a user-specified window length $winlen > 0$, the \underline{problem} of mining maximal frequent episodes is to enumerate all frequent episodes that are not strictly included in another frequent episode~\cite{10.1016/j.is.2007.07.003}. An episode $\alpha = \langle Y_1, Y_2, \ldots , Y_i\rangle$ is \underline{strictly included} in an episode $\beta = \langle X_1, X_2, \ldots , X_p\rangle$ if and only if $Y_1 \subseteq X_{k_1}, Y_2 \subseteq X_{k_2}, \ldots, Y_i \subseteq X_{k_i}$ for some integers $1 \le k_1 < k_2 < \ldots < k_i \le p$. This relation is denoted as $ \alpha \sqsubseteq \beta$.
\end{definition}

\begin{definition}[Location list]
    Let there be an input sequence $S = \langle (SE_{t_1}, t_1), (SE_{t_2}, t_2), \ldots, (SE_{t_n}, t_n) \rangle$. Furthermore, assume that events from each event set in $S$ are sorted by a total order $ \prec$ on events. This order can be any order such as the lexicographical order ($a \prec b \prec c \prec d$). If an event $e$ is included in the $i$-th event set $SE_{t_i}$ of the input sequence, then $e$ is said to appear at position $P_w = \sum_{w=1, \ldots, i-1} |SE_{t_w}| + |\{y|y \in SE_{t_i} \land y \prec e\}|$. For an event $e$, its location list in the sequence $S$ is the list of its timestamps and is denoted as $locList(e)$. An interesting property is that the support of an event $e$ can be obtained from its location list as $sup(e) = |locList(e)|$.
\end{definition}

\begin{definition}[Bound list]
    Let there be a re-encoded sequence $S' = \langle(SE_{t_1}, t_1), (SE_{t_2}, t_2), \ldots, (SE_{t_n}, t_n)\rangle$. The bound list of a parallel episode $pe$ is defined as $boundList(pe) = \{[t, t]|pe \subseteq SE_t \in S'\}$. The bound list of the serial extension of a composite episode $ep$ with $pe$, is defined as: $boundList(sExtension(ep, pe)) =
\{[u, w]| [u, v] \in boundList(ep) \land [w, w] \in boundList(pe) \land w - u < winlen
\land v < w\}$. The bound list of a composite episode $ep$ allows deriving its support as
$sup(ep) = |\{t_s|[t_s, t_e] \in boundList(ep)\}|$.
\end{definition}

\section{Related Work}\label{sec:relwork}

Frequent episode mining is an active research area in data mining that has been extensively studied over the past 30 years~\cite{https://doi.org/10.1002/widm.1524}. The task was first introduced by Mannila et al.~\cite{10.1023/A:1009748302351, 10.5555/3001460.3001490}, who proposed the WINEPI and MINEPI algorithms for discovering frequent episodes in event sequences. Since then, numerous algorithms have been developed to address various aspects of this problem. Episode mining has found applications in numerous domains, including web stream analysis, log analysis, network fault management, cybersecurity, and prediction~\cite{https://doi.org/10.1002/widm.1524}.

Frequent episode mining algorithms can be categorized based on their search strategy into breadth-first and depth-first approaches. Breadth-first algorithms, such as WINEPI and MINEPI~\cite{10.1023/A:1009748302351, 10.5555/3001460.3001490}, alternate between candidate generation and frequency checking phases. Depth-first algorithms, such as EMMA~\cite{10.1016/j.is.2007.07.003}, recursively extend episodes and achieve better memory efficiency. Various frequency definitions have been proposed for counting episode occurrences, including window-based frequency, minimal occurrence-based frequency, head frequency, and non-overlapped occurrence-based frequency~\cite{https://doi.org/10.1002/widm.1524}. The head frequency measure, which counts distinct start points of occurrences, has been adopted by several recent algorithms including EMMA~\cite{10.1016/j.is.2007.07.003} due to its suitability for prediction tasks. The choice of frequency definition significantly impacts both the set of discovered episodes and the computational complexity of the mining process.

Several extensions of traditional episode mining have been proposed~\cite{https://doi.org/10.1002/widm.1524}: top-k episode mining allows users to specify the number of episodes to discover; constraint-based episode mining integrates gap or span constraints; high utility episode mining considers the utility of events rather than just frequency; online and stream episode mining is designed for scenarios where the event sequence is continuously updated; weighted episode mining, where weight is assigned to each event type to reflect differing importance; and many others.

In addition to these extensions, concise representations of frequent episodes also represent an important extension in episode mining. They address a major challenge in frequent episode mining -- the potentially huge number of discovered patterns, which can be difficult for users to analyze~\cite{10.1007/978-3-031-20992-5-8}. To handle this problem, researchers have suggested mining concise representations that summarize the complete set of frequent episodes. According to~\cite{https://doi.org/10.1002/widm.1524}, closed episodes are frequent episodes that have no proper super-episode with same support count, and maximal episodes are frequent episodes that have no frequent super-episode, providing an even more compact representation. Two algorithms have been proposed for mining maximal episodes: LA-FEMH+~\cite{10.1145/3326163}, which is restricted to serial episodes only, and MaxFEM~\cite{10.1007/978-3-031-20992-5-8}, which can discover both serial and parallel episodes in complex event sequences.

Episode mining is related to but distinct from other pattern mining tasks. Sequential pattern mining aims to find patterns common to multiple sequences, while episode mining focuses on a single long sequence~\cite{https://doi.org/10.1002/widm.1524}. Recent work on a sequential rule mining with gap constraints~\cite{10.1109/TKDE.2023.3241213} addresses a related problem of mining maximal co-occurrence rules, but operates on simple sequences without simultaneous events.
Similar challenges related to the exponential growth of candidate patterns also appear in other pattern mining areas. In the context of high average utility itemset mining, Nandhini and Kannimuthu~\cite{10.3233/JIFS-231852} propose an optimization-based approach that formulates the mining process as a search problem guided by utility-based evaluation functions and adaptive thresholding. Instead of using exhaustive enumeration, their method applies a metaheuristic strategy to reduce the search space and limit the number of generated candidates.
This shows that controlling the result size and pattern relevance, achieved through maximality constraints or through optimization-driven selection, is a widely applicable idea across several pattern mining domains and a fundamental principle for designing scalable mining algorithms.

The MaxFEM algorithm~\cite{10.1007/978-3-031-20992-5-8} represents the state-of-the-art for mining maximal frequent episodes in complex event sequences with simultaneous events. It is based on the search procedure originally introduced in EMMA~\cite{10.1016/j.is.2007.07.003} and employs three optimization strategies: Efficient Filtering of Non-maximal episodes (EFE), Skip Extension checking (SEC), and  Temporal pruning (TP). The experimental evaluation showed that MaxFEM achieves 10--40\% speedup over EMMA while significantly reducing the number of output patterns. However, despite these optimizations, there remain opportunities for further performance improvements, particularly in terms of computational efficiency on large-scale event sequences. In this work, we focus on optimizing the MaxFEM algorithm through parallelization of its core search procedure.

\section{MaxFEM Algorithm}\label{sec:maxfem}

The MaxFEM (Maximal Frequent Episode Miner) algorithm is designed to discover maximal frequent episodes in complex event sequences. Unlike traditional frequent episode mining approaches that generate redundant patterns, MaxFEM focuses on identifying only those frequent episodes that are not strictly included in any other frequent episode. This section provides a description of the algorithm's main stages based on the original publication~\cite{10.1007/978-3-031-20992-5-8}.

\begin{algorithm2e}
\caption{The MaxFEM algorithm}
\SetKwInOut{Input}{input}
\SetKwInOut{Output}{output}

\Input{$S$: an input sequence, $minsup$: a user-specified threshold, $winlen$: the window length}
\Output{the maximal frequent episodes}

Scan $S$ to calculate $sup(e)$ for each event $e \in E$;

$E' \leftarrow \{e|e \in E \land sup(e) \geq minsup\}$;

Thereafter, ignore or remove each event $e \notin E'$ from $S$;

Read $S$ to build the location list of each frequent event $e \in E'$;

$PEpisodes \leftarrow E'$;

\ForEach{\emph{parallel episode} $ep \in PEpisodes$ \emph{such that} $sup(ep) \geq minsup$}{
    \ForEach{\emph{event} $e \in E'$ \emph{such that} $sup(e) \geq minsup$}{
        $newE \leftarrow parallelExtension(ep, e)$; \texttt{// and build} $newE$\texttt{'s location list}

        \If{$sup(newE) \geq minsup$}{
            $PEpisodes \leftarrow PEpisodes \cup \{newE\}$;
        }
    }
}
$W \leftarrow PEpisodes$;

Re-encode the sequence $S$ into a sequence $S'$ using the parallel episodes;

\ForEach{\emph{composite episode} $ep \in W$ \emph{such that} $sup(ep) \geq minsup$}{
    \ForEach{\emph{event} $e \in PEpisodes$ \emph{such that} $sup(e) \geq minsup$}{
        $newE \leftarrow sExtension(ep, e)$; \texttt{// and build} $newE$\texttt{'s bound list}

        \If{$sup(newE) \geq minsup$ \emph{and} $newE$ \emph{has no superset in} $W$}{
            $W \leftarrow W \cup \{newE\}$;

            Remove all subsets of $newE$ that are in $W$;
        }
    }
}
Return $W$;
\end{algorithm2e}

\subsection{Algorithm Overview}

The MaxFEM algorithm receives as input a complex event sequence $S$, a minimum support threshold $minsup$, and a maximum window length $winlen$. The output consists of the set of maximal frequent episodes.

The main idea of the algorithm is to use a depth-first search strategy to explore the space of frequent episodes while keeping a set $W$ that stores the current maximal frequent episodes. When a new frequent episode $\alpha$ is found, the algorithm checks if it is maximal: if $\alpha$ is strictly included in any episode already in $W$, it is discarded; otherwise, $\alpha$ is added to $W$, and all episodes strictly included in $\alpha$ are removed from the set $W$.

\subsection{Algorithm steps}

Step 1: Finding the frequent events. In this stage, the algorithm scans the input sequence $S$ to compute the support of each event. The set $E'$ of frequent events that satisfy the condition $sup(e) \geq minsup$ is formed. Infrequent events are excluded from further processing.

Step 2: Building the location lists. For each frequent event, a data structure called a location list is built. The location list of an event $e$ contains the timestamps of all occurrences of that event in the sequence and is denoted as $locList(e)$. The support of an event can be computed as the size of its location list: $sup(e) = |locList(e)|$.

Step 3: Finding the frequent parallel episodes. In this step, frequent events are extended recursively to form parallel episodes. A parallel extension of an episode $ep$ with an event $e$ creates a new episode $newE = ep \cup \{e\}$. The location list of the new episode is computed as the intersection of the location lists of the original episode and the added event: $locList(newE) = locList(ep) \cap locList(e)$. If the support of the new episode is at least $minsup$, it is added to the set of frequent parallel episodes $PEpisodes$.

Step 4: Sequence re-encoding. Each frequent parallel episode is assigned a unique identifier. The original sequence $S$ is transformed into a re-encoded sequence $S'$ by replacing events with their corresponding parallel episodes.

Step 5: Finding the maximal frequent composite episodes. In the final step, the algorithm searches for composite episodes using serial extensions. A serial extension of an episode $ep = \langle SE_1, SE_2, \ldots, SE_x \rangle$ with a parallel episode $pe$ creates a composite episode $sExtension(ep, pe) = \langle SE_1, SE_2, \ldots, SE_x, pe \rangle$.

To compute the support of composite episodes, the algorithm uses bound lists. The support is calculated as the number of distinct starting points in the bound list. When a new frequent composite episode is found, the algorithm checks if it is maximal by comparing it with the episodes in the set $W$. Episodes that are not maximal are removed from the result set.

\subsection{Running example}

To illustrate how the MaxFEM algorithm works, consider the complex event sequence shown in Fig. ~\ref{fig:example-sequence} with $minsup = 2$ and $winlen = 2$.

\begin{figure}[h]
\centering
\begin{tikzpicture}[>=stealth, node distance=1cm]
    \node[draw, minimum width=0.7cm, minimum height=0.7cm] (t1) {$a, c$};
    \node[draw, minimum width=0.7cm, minimum height=0.7cm, right of=t1] (t2) {$a$};
    \node[draw, minimum width=0.7cm, minimum height=0.7cm, right of=t2] (t3) {$a, b$};
    \node[draw, minimum width=0.7cm, minimum height=0.7cm, right of=t3] (t6) {$a$};
    \node[draw, minimum width=0.7cm, minimum height=0.7cm, right of=t6] (t7) {$a, b$};
    \node[draw, minimum width=0.7cm, minimum height=0.7cm, right of=t7] (t8) {$c$};
    \node[draw, minimum width=0.7cm, minimum height=0.7cm, right of=t8] (t9) {$b$};
    \node[draw, minimum width=0.7cm, minimum height=0.7cm, right of=t9] (t11) {$d$};

    \node[below=0.3cm of t1] {$t_1$};
    \node[below=0.3cm of t2] {$t_2$};
    \node[below=0.3cm of t3] {$t_3$};
    \node[below=0.3cm of t6] {$t_6$};
    \node[below=0.3cm of t7] {$t_7$};
    \node[below=0.3cm of t8] {$t_8$};
    \node[below=0.3cm of t9] {$t_9$};
    \node[below=0.3cm of t11] {$t_{11}$};
\end{tikzpicture}
\caption{Input complex event sequence $S$}
\label{fig:example-sequence}
\end{figure}
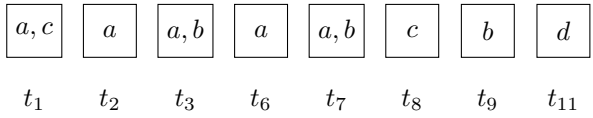

\begin{figure}[h]
\centering
\resizebox{\columnwidth}{!}{%
\begin{tikzpicture}[
    box/.style={draw, rounded corners, minimum width=2.2cm, minimum height=0.8cm, align=center, font=\scriptsize},
    arrow/.style={->, thick, >=stealth},
    label/.style={font=\tiny, align=left}
]

\node[box, fill=blue!10] (step1) {
    \textbf{Step 1}\\
    Find Frequent Events
};
\node[label, right=0.2cm of step1] (step1-result) {
    $sup(a) = 5, sup(b) = 3$\\
    $sup(c) = 2, sup(d) = 1$\\
    $E' = \{a, b, c\}$
};

\node[box, fill=blue!10, below=0.6cm of step1] (step2) {
    \textbf{Step 2}\\
    Build Location Lists
};
\node[label, right=0.2cm of step2] (step2-result) {
    $locList(a) = \{t_1, t_2, t_3, t_6, t_7\}$\\
    $locList(b) = \{t_3, t_7, t_9\}$\\
    $locList(c) = \{t_1, t_8\}$
};

\node[box, fill=blue!10, below=0.6cm of step2] (step3) {
    \textbf{Step 3}\\
    Find Parallel Episodes
};
\node[label, right=0.2cm of step3] (step3-result) {
    $PEpisodes = \{\langle\{a\}\rangle, \langle\{b\}\rangle,$\\
    $\langle\{c\}\rangle, \langle\{a,b\}\rangle\}$\\
    $sup(\{a,b\}) = 2 \geq minsup$
};

\node[box, fill=blue!10, below=0.6cm of step3] (step4) {
    \textbf{Step 4}\\
    Re-encode Sequence
};
\node[label, right=0.2cm of step4] (step4-result) {
    $\#1 = \langle\{a\}\rangle, \#2 = \langle\{b\}\rangle$\\
    $\#3 = \langle\{c\}\rangle, \#4 = \langle\{a,b\}\rangle$
};

\node[box, fill=blue!10, below=0.6cm of step4] (step5) {
    \textbf{Step 5}\\
    Find Maximal Episodes
};
\node[label, right=0.2cm of step5] (step5-result) {
    $\langle\{a\}, \{a,b\}\rangle$: $sup = 2$\\
    Remove: $\langle\{a\}\rangle, \langle\{b\}\rangle, \langle\{a,b\}\rangle$
};

\node[box, fill=green!20, below=0.6cm of step5] (output) {
    \textbf{Output}
};
\node[label, right=0.2cm of output] (output-result) {
    $W = \{\langle\{c\} \rangle, \langle\{a\}, \{a,b\}\rangle\}$
};

\draw[arrow] (step1) -- (step2);
\draw[arrow] (step2) -- (step3);
\draw[arrow] (step3) -- (step4);
\draw[arrow] (step4) -- (step5);
\draw[arrow] (step5) -- (output);

\end{tikzpicture}%
}
\caption{Step-by-step execution of the MaxFEM algorithm}
\label{fig:maxfem-example}
\end{figure}
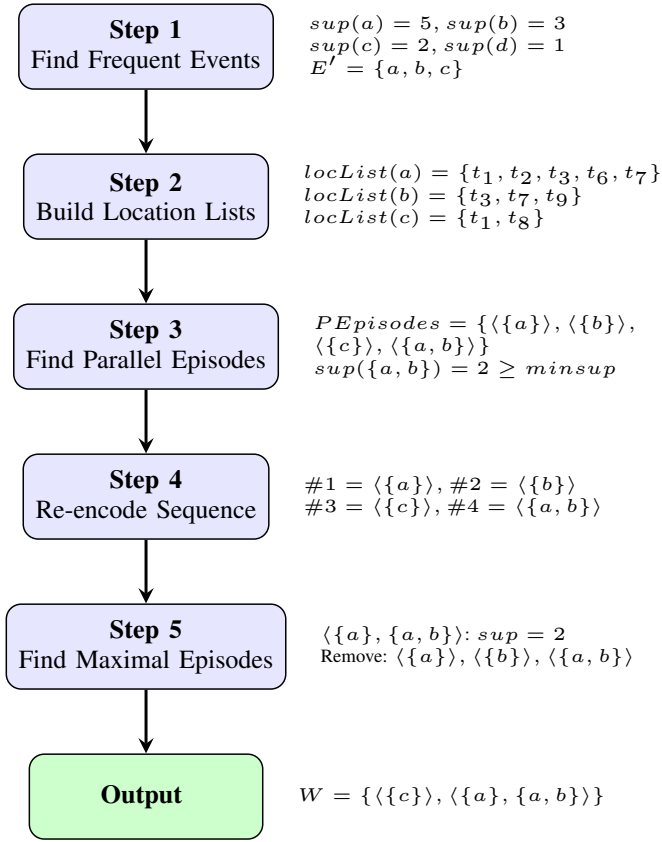

Fig. ~\ref{fig:maxfem-example} shows how MaxFEM works on the input sequence step by step.

In Step 1, the algorithm scans the sequence and counts how many times each event appears. Events $a$, $b$, and $c$ are frequent because their support values (5, 3, and 2) are at least $minsup = 2$. Event $d$ appears only once, so it is not frequent and is ignored.

In Step 2, the algorithm builds a location list for each frequent event. For example, event $a$ appears at timestamps $t_1, t_2, t_3, t_6, t_7$, so $locList(a) = \{t_1, t_2, t_3, t_6, t_7\}$.

In Step 3, the algorithm tries to combine frequent events to form parallel episodes. It computes the location list of each combination as the intersection of location lists. For example, $locList(\{a,b\}) = locList(a) \cap locList(b) = \{t_3, t_7\}$, so $sup(\{a,b\}) = 2$. This episode is frequent. However, $\langle\{a,c\}\rangle$ has $sup = 1$, so it is not frequent. After this stage, $PEpisodes = \{\langle\{a\}\rangle, \langle\{b\}\rangle, \langle\{c\}\rangle, \langle\{a,b\}\rangle\}$.

In Step 4, each parallel episode gets a unique identifier ($\#1$, $\#2$, $\#3$, $\#4$), and the sequence is re-encoded using these identifiers.

In Step 5, the algorithm looks for composite episodes by performing serial extensions. For example, $\langle\{a\}\rangle$ is extended to $\langle\{a\}, \{a,b\}\rangle$, which has $sup = 2$ and is frequent. Now the algorithm checks maximality: since $\langle\{a\}\rangle$, $\langle\{b\}\rangle$, and $\langle\{a,b\}\rangle$ are all strictly included in $\langle\{a\}, \{a,b\}\rangle$, they are removed from $W$.

The final output contains two maximal frequent episodes: $\langle\{c\}\rangle$ and $\langle\{a\}, \{a,b\}\rangle$.

\subsection{Optimization Strategies}

The original MaxFEM algorithm uses three strategies to improve performance.

Strategy 1. Efficient Filtering of Non-maximal episodes (EFE). The set $W$ is implemented as a collection of heaps $W = \{W_1, W_2, \ldots, W_n\}$, where $W_x$ contains maximal episodes of size $x$. To check if an episode of size $w$ is included in another episode, it is enough to compare it only with episodes of corresponding sizes. Also, episode hashing is used to speed up comparisons.

Strategy 2. Skip Extension checking (SEC). If a frequent episode is successfully extended by a serial extension, the maximality check is skipped, because such an episode cannot be maximal.

Strategy 3. Temporal pruning (TP). When building a bound list for an extension, if the number of remaining elements is not enough to reach the threshold $minsup$, the construction is stopped early.

\section{Improvements}\label{sec:improvements}

While the original MaxFEM algorithm~\cite{10.1007/978-3-031-20992-5-8} demonstrates good performance compared to EMMA, our implementation introduces several improvements that significantly enhance its efficiency and scalability. This section describes key optimizations applied to the algorithm.

\subsection{Performance Benefits of C++ Implementation}

The first major improvement is the reimplementation of the algorithm in C++. While the original MaxFEM is implemented in Java as part of the SPMF library, our implementation uses C++ to achieve better runtime performance and lower memory consumption. C++ provides direct memory management and eliminates runtime overhead such as garbage collection pauses, which is particularly beneficial for memory-intensive algorithms like frequent episode mining. Additionally, C++ allows fine-grained control over data layout and cache efficiency, which becomes crucial when processing large event sequences with millions of timestamps.

\subsection{Memory-Efficient Data Structures}

Our implementation applies several memory optimization techniques throughout all algorithm steps:

\textbf{Pre-allocation and move semantics.} When the required size of a data structure is known in advance, we use \texttt{reserve()} to allocate memory once, avoiding multiple reallocations. We also use C++ move semantics extensively to transfer ownership of large data structures without copying. This is particularly important when building new episodes and extending bound lists, where data can be moved rather than copied.

\textbf{Shared storage for location lists.} Location lists for parallel episodes are stored using shared pointers \texttt{std::shared\_ptr}. This allows multiple episodes that contain the same events to reference the same location list without duplication, reducing memory consumption when many episodes share event locations.

These techniques work together to reduce both peak memory usage and the number of memory allocations, which directly translates to improved cache behavior and reduced memory allocation overhead.

\subsection{Efficient Implementation of Steps 1-4}

The first four steps of MaxFEM -- finding frequent events, building location lists, discovering parallel episodes, and re-encoding the sequence -- primarily benefit from efficient data structure design and careful implementation.

In Step 1, after counting event occurrences, frequent events are mapped to a contiguous range of integers $[0, |E'| - 1]$, while infrequent events are removed. This mapping allows the use \texttt{std::vector} instead of \texttt{std::map} or \texttt{std::unordered\_map} for storing events metadata. Consequently, access time is reduced to $O(1)$ with lower constant factors, and cache locality is significantly improved.

In Step 2, location lists are built in a single pass over the input sequence. The lists are sorted as vectors of timestamps with pre-allocated capacity based on event support, minimizing reallocations.

In Step 3, frequent parallel episodes are discovered by recursive depth-first search. The parallel extension operation creates new episodes by computing the intersection of location lists. This intersection takes advantage of the sorted nature of location lists, resulting in linear-time complexity $O(n + m)$ where $n$ and $m$ are the list sizes. The use of shared pointers for location lists becomes particularly beneficial here, as episodes with identical event sets reference the same underlying storage.

Step 4 is skipped entirely. Explicit creation of re-encoded sequence $S'$ was observed to be redundant because the event sequence is no longer required once the location lists are built in Step 2. Instead, the parallel episodes identified in Step 3 serve directly as the basis for the serial extension process in Step 5.

These stages execute sequentially, but their efficient implementation ensures they consume only a small part of total runtime compared to Step 5.

\subsection{Parallelization of Composite Episode Mining}

Step 5 is the most computationally intensive part of the algorithm, as it performs a depth-first search through the space of composite episodes. To improve its scalability, we introduce task-based parallelization using thread pools with dynamic load balancing.

The parallelization strategy is based on the following observation: each seed parallel episode defines an independent subtree in the search space of composite episodes. Therefore, the search can be naturally decomposed into independent tasks, one for each seed. We use the Boost.Asio thread pool to distribute these tasks across multiple CPU cores. The thread pool employs an efficient work distribution mechanism similar to work stealing, where idle threads can pick up tasks from a shared queue.

To achieve better load balancing, we implement adaptive dynamic task splitting. When processing a composite episode, the algorithm attempts to spawn a new parallel task for each serial extension if the number of currently running tasks is below a specified threshold. This threshold is computed as the product of the number of threads and a user-configurable multiplier (by default, 1). If the threshold is exceeded, the current task continues exploring the subtree sequentially without spawning new tasks. This strategy balances parallelism and overhead: when threads would be idle, new tasks are spawned to keep them busy; when sufficient work exists, sequential processing avoids the cost of excessive task creation.

Each parallel task maintains a local buffer for storing maximal episodes discovered within its subtree. This design minimizes synchronization overhead, as threads do not need to acquire locks during the search process. Once a task is finished, it moves its local buffer to the global result collection under a mutex lock. After all tasks complete, the global collection performs a batch merge of local results using the Efficient Filtering of Non-maximal episodes strategy from the original MaxFEM paper~\cite{10.1007/978-3-031-20992-5-8}. Episodes are processed in descending order of length and grouped by size, with hash-based comparisons used to quickly reject candidates before performing full subset checks. The processing order eliminates the need for sub-episode checks and prevents the insertion and subsequent removal of non-maximal episodes, as candidates are only checked against already confirmed larger maximal episodes. While this approach results in a minor memory overhead due to the temporary storage of some non-maximal episodes, it enables high scalability by eliminating lock contention during the parallel search.

The combination of efficient C++ implementation, careful memory management, and task-based parallelism results in significant performance improvements over the original Java implementation, as demonstrated in Section~\ref{sec:eval}.

\subsection{Python integration}

Python has established itself as the de facto standard language in the Data Science and Machine Learning domains. Consequently, providing seamless integration with the Python ecosystem is crucial for the adoption of any modern data profiling tool. Desbordante addresses this need by offering a native Python package distributed via PyPI, which can also be built locally from sources.

This approach contrasts significantly with SPMF, where the primary usage pattern involves executing a Java Virtual Machine (JVM) process via the command line. While SPMF can be invoked from Python using the \texttt{subprocess} mechanism, this method introduces several critical drawbacks:
\begin{enumerate}
\item \textbf{Dependency Management.} The user is required to have a pre-installed and compatible JVM configured in the environment.

\item \textbf{Engineering Overhead.} Users are forced to implement custom ``glue code'' to transform their data into the rigid SPMF text format (e.g., serializing events, handling specific delimiters). This manual serialization and subsequent result parsing increase code complexity and the risk of formatting errors.

\item \textbf{I/O and Memory Bottlenecks.} The necessity of writing intermediate files creates significant disk I/O overhead. Furthermore, serializing large datasets from memory to disk and reading them back doubles the storage requirements during execution.

\item \textbf{Lack of Interactivity.} The workflow prevents the use of in-memory data structures, making integration with interactive environments (like Jupyter Notebooks) cumbersome.
\end{enumerate}

Desbordante overcomes these limitations by exposing a native Python interface. Our implementation allows passing a Complex Event Sequence directly as a Python \texttt{Iterable} object, where each element represents an Event Set (optionally paired with a timestamp in a tuple).

This design enables a streamlined pipeline where data preprocessing, mining, and post-processing occur within a single memory space. Unlike MaxFEM, which requires the full dataset to be loaded into memory, our approach allows users to leverage Python generators to populate the algorithm's internal C++ structures directly. This eliminates the need to materialize the entire dataset as heavy Python objects or write custom format converters. By removing intermediate disk I/O and reducing the engineering effort required for data preparation, Desbordante facilitates smoother integration of sequence mining into analytical workflows.

\subsection{Usage Example}

This code (Listing~\ref{lst:parmaxfem_usage}) illustrates the practical usage of ParMaxFEM through Desbordante's Python interface.

\begin{lstlisting}[style=pythonstyle, caption={ParMaxFEM Usage Example}, label={lst:parmaxfem_usage}]
import desbordante as db

# Input sequence (events, timestamp)
seq = [({1, 3}, 1), ({1}, 2),
       ({1, 2}, 3), ({1}, 6),
       ({1, 2}, 7), ({3}, 8),
       ({2}, 9), ({4}, 11)]

algo = db.fem.ParMaxFEM()
algo.load_data(sequence=seq)

# Alternatively, load from file in SPMF format:
# algo.load_data("sequence.txt")

algo.execute(minsup=2, window_size=2)

episodes = algo.get_max_frequent_episodes()
print(f"Found {len(episodes)} maximal episodes")
for ep, sup in episodes:
    print(f"Episode: {ep}, Support: {sup}")

# Output:
# Found 2 maximal episodes
# Episode: [[3]], Support: 2
# Episode: [[1], [1, 2]], Support: 2
\end{lstlisting}

This example demonstrates the key advantages of Desbordante’s Python integration: event sequences and resulting
episodes are defined using native Python data structures (lists,
sets and tuples) and the API is simple and clear.

\section{Evaluation}\label{sec:eval}

\subsection{Methodology and Configuration}

To ensure rigorous performance analysis and reproducibility, particularly for the multi-threaded implementation, we used a specific hardware and software configuration. Most experiments were conducted on a laptop powered by an Intel\textsuperscript{\textregistered} Core\textsuperscript{\texttrademark} Ultra 7 165U processor (Meteor Lake architecture) with 32 GiB of RAM, running the Linux\textsuperscript{\textregistered} operating system.

Given the heterogeneous architecture of the CPU, we employed CPU shielding using the \texttt{tuna} tool to isolate the benchmarking environment. The operating system and background processes were restricted to the 2 Performance (P) cores and 2 Low-Power Efficient (LP-E) cores. The benchmarks were executed exclusively on the 8 isolated Efficient (E) cores, locked at a fixed frequency of 3.8 GHz.

Hardware power-saving features such as Intel\textsuperscript{\textregistered} SpeedStep\textsuperscript{\textregistered} and Intel\textsuperscript{\textregistered} Turbo Boost\textsuperscript{\texttrademark} were disabled via BIOS settings to prevent frequency scaling. The Linux CPU governor was explicitly set to \texttt{performance} using the \texttt{cpupower} tool. To further minimize OS jitter, we configured the kernel boot parameters (via GRUB) with \texttt{nohz\_full} (full tickless mode) and \texttt{rcu\_nocbs} (offloaded RCU callbacks) for isolated cores. Additionally, deep sleep states were disabled (\texttt{intel\_idle.max\_cstate=0}, \texttt{processor.max\_cstate=1}) to eliminate wakeup latency.

The code was executed within a Docker\textsuperscript{\textregistered} container based on Ubuntu 24.04 with a hard memory limit of 30 GiB and a disabled swap. To avoid OverlayFS overhead, input/output directories were mounted using bind mounts (\texttt{-v} option). Our C++ implementations were compiled using \texttt{GCC 13.2.0} with the \texttt{-O3} optimization flag (Release build type). The Java-based algorithm (SPMF) was executed on the \texttt{OpenJDK}\textsuperscript{\textregistered} \texttt{21} Runtime Environment (LTS). Execution time and peak memory usage were measured using the \texttt{GNU Time} utility.

To evaluate the parallel scalability of ParMaxFEM under higher thread counts (Experiment 4), we utilized an additional cloud-based hardware configuration. This specific experiment was conducted on a virtual machine powered by a 2.0 GHz Intel\textsuperscript{\textregistered} Xeon\textsuperscript{\textregistered} Gold 6338 processor with 32 physical cores, providing 64 vCPUs and 40 GiB of RAM.

\subsection{Research Questions}

To systematically evaluate the performance improvements and scalability of our proposed solution, we designed a comprehensive experimental study. Our evaluation aims to verify the benefits of both the system-level optimizations inherent to the C++ implementation and the algorithmic advantages of the parallelization. Specifically, this study addresses the following three research questions:
\begin{enumerate}
\item \textbf{Implementation Efficiency (RQ1).} How much gain in terms of both execution speed and memory usage is achieved solely by re-implementing the MaxFEM algorithm in C++ with system-level optimizations compared to the existing Java implementation (SPMF)?
\item \textbf{Performance Superiority (RQ2).} To what extent does the proposed parallel algorithm (ParMaxFEM) outperform both the baseline Java implementation (SPMF) and the optimized sequential C++ version, and what is the overall speedup achieved across diverse datasets?
\item \textbf{Parallel Scalability (RQ3).} How effectively does the proposed algorithm scale as the number of execution threads increases, and at what point does it reach its scalability limit?
\end{enumerate}

By answering these questions, we aim to quantify the impact of our architectural decisions and demonstrate the practical viability of using Desbordante ParMaxFEM for mining maximal frequent episodes.

\begin{table}[ht]
\caption{Datasets}
\label{tab:datasets}
\centering
\resizebox{\columnwidth}{!}{
\begin{tabular}{l r r r r l}
\hline
\textbf{Datasets} & \textbf{Trans.} & \textbf{Items} & \textbf{Avg. Len.} & \textbf{Dens. (\%)} & \textbf{Type} \\
\hline
accidents & 340,183 & 468 & 33.80 & 7.22 & Traffic \\
BMS1 & 59,602 & 497 & 2.51 & 0.51 & Click-stream \\
BMS2 & 77,512 & 3,340 & 4.62 & 0.14 & Click-stream \\
chainstore & 1,112,949 & 46,086 & 7.23 & 0.02 & Grocery \\
chess & 3,196 & 75 & 37.00 & 49.33 & Game steps \\
kosarak & 990,002 & 41,270 & 8.10 & 0.02 & Click-stream \\
mushrooms & 8,416 & 119 & 23.00 & 19.33 & Biological \\
OnlineRetail & 541,909 & 2,603 & 4.37 & 0.17 & E-commerce \\
PAMAP & 1,000,000 & 141 & 23.93 & 16.97 & Sensors \\
RecordLink & 574,913 & 29 & 10.00 & 34.48 & Linkage \\
retail & 88,162 & 16,470 & 10.30 & 0.06 & Retail \\
Skin & 245,057 & 11 & 4.00 & 36.36 & Segmentation \\
\hline
\multicolumn{6}{l}{\footnotesize \textit{Trans.}: Transaction count, \textit{Dens.}: Density ($\frac{\text{Avg.Len}}{\text{Items}} \times 100$).}
\end{tabular}
}
\end{table}

\begin{figure*}
    \centering
    \includegraphics[width=1\linewidth]{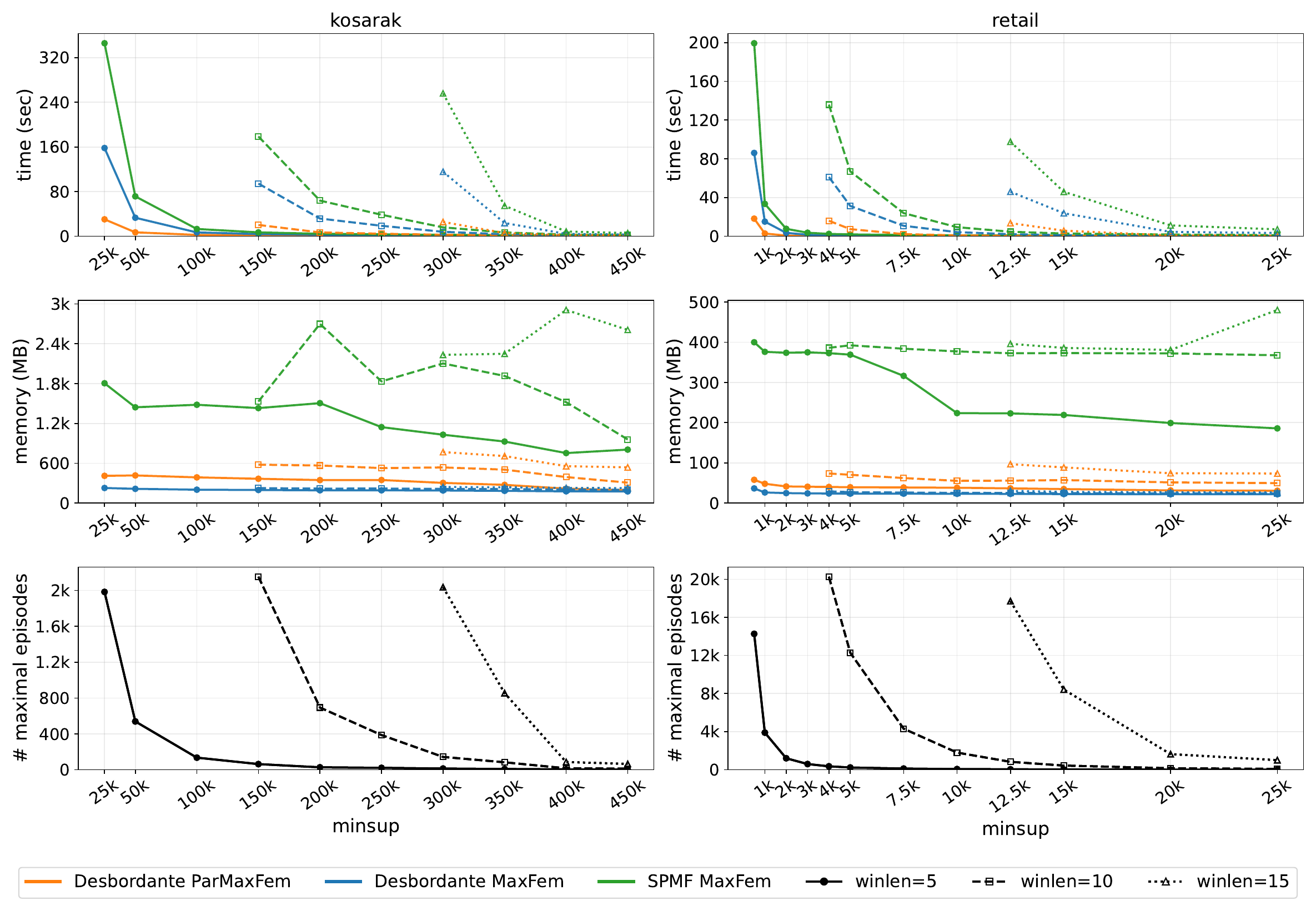}
    \caption{Comparison of MaxFEM implementations: runtime, memory usage, and number of maximal episodes on Kosarak (left) and Retail (right) datasets for different minimum support values and window lengths.}
    \label{fig:kosarak-retail}
\end{figure*}

\begin{table*}[h]
\caption{Comparative Benchmark Results: Desbordante vs. SPMF}
\label{tab:perf_final_fixed_data}
\centering
\scriptsize
\setlength{\tabcolsep}{2pt}
\renewcommand{\arraystretch}{1.2}

\begin{tabular}{
    l
    l
    >{\columncolor{blue!5}\centering\arraybackslash}p{0.11\textwidth}
    >{\columncolor{red!5}\centering\arraybackslash}p{0.11\textwidth}
    >{\columncolor{blue!5}\centering\arraybackslash}p{0.11\textwidth}
    >{\columncolor{red!5}\centering\arraybackslash}p{0.11\textwidth}
    >{\columncolor{blue!5}\centering\arraybackslash}p{0.11\textwidth}
    >{\columncolor{red!5}\centering\arraybackslash}p{0.11\textwidth}
}
\toprule
& & \multicolumn{2}{c}{\textbf{Int. 1: Small} ($2$s -- $1$m)} & \multicolumn{2}{c}{\textbf{Int. 2: Medium} ($1-5$m)} & \multicolumn{2}{c}{\textbf{Int. 3: Large} ($5-10$m)} \\
\cmidrule(lr){3-4} \cmidrule(lr){5-6} \cmidrule(lr){7-8}
\textbf{Dataset} & \textbf{Algorithm} & \textbf{Time (Spd)} & \textbf{Mem (Fac)} & \textbf{Time (Spd)} & \textbf{Mem (Fac)} & \textbf{Time (Spd)} & \textbf{Mem (Fac)} \\
\midrule

\multirow{3}{*}{\textbf{accidents}}
& SPMF MaxFEM & 21.44 / 1.00 & 1720.4 / 1.00 & 153.42 / 1.00 & 1802.7 / 1.00 & 580.78 / 1.00 & 2012.7 / 1.00 \\
& Desbordante MaxFEM & 6.42 / 3.06 & \textbf{207.9 / 8.30} & 41.62 / 3.67 & \textbf{215.4 / 8.40} & 118.98 / 4.88 & \textbf{217.1 / 9.27} \\
& \textbf{Desbordante ParMaxFEM} & \textbf{2.22 / 8.26} & 569.4 / 3.24 & \textbf{9.14 / 16.05} & 858.5 / 2.29 & \textbf{30.95 / 18.77} & 1620.5 / 1.24 \\
\cmidrule{1-8}

\multirow{3}{*}{\textbf{BMS1}}
& SPMF MaxFEM & 17.54 / 1.00 & 369.9 / 1.00 & 113.75 / 1.00 & 434.9 / 1.00 & 390.41 / 1.00 & 542.3 / 1.00 \\
& Desbordante MaxFEM & 10.13 / 1.90 & \textbf{26.1 / 16.36} & 82.41 / 1.44 & \textbf{63.2 / 7.34} & 263.02 / 1.53 & \textbf{116.0 / 5.58} \\
& \textbf{Desbordante ParMaxFEM} & \textbf{3.19 / 6.59} & 29.0 / 14.59 & \textbf{25.18 / 5.49} & 70.7 / 6.39 & \textbf{85.68 / 6.64} & 133.2 / 4.50 \\
\cmidrule{1-8}

\multirow{3}{*}{\textbf{BMS2}}
& SPMF MaxFEM & 12.60 / 1.00 & 427.2 / 1.00 & 160.42 / 1.00 & 614.5 / 1.00 & 387.69 / 1.00 & 702.5 / 1.00 \\
& Desbordante MaxFEM & 7.14 / 1.76 & \textbf{31.5 / 13.70} & 88.16 / 1.87 & \textbf{97.7 / 7.09} & 182.82 / 2.14 & \textbf{140.4 / 5.35} \\
& \textbf{Desbordante ParMaxFEM} & \textbf{2.50 / 4.99} & 32.7 / 13.18 & \textbf{20.61 / 7.74} & 102.1 / 6.73 & \textbf{40.20 / 9.71} & 148.4 / 4.99 \\
\cmidrule{1-8}

\multirow{3}{*}{\textbf{chainstore}}
& SPMF MaxFEM & 11.12 / 1.00 & 934.5 / 1.00 & 126.13 / 1.00 & 1542.0 / 1.00 & 423.28 / 1.00 & 1440.9 / 1.00 \\
& Desbordante MaxFEM & 7.45 / 1.40 & \textbf{166.2 / 5.61} & 60.92 / 2.15 & \textbf{203.5 / 7.69} & 208.94 / 2.04 & \textbf{260.5 / 5.77} \\
& \textbf{Desbordante ParMaxFEM} & \textbf{4.08 / 2.44} & 171.6 / 5.40 & \textbf{18.00 / 7.49} & 219.2 / 7.13 & \textbf{58.07 / 8.14} & 304.2 / 4.83 \\
\cmidrule{1-8}

\multirow{3}{*}{\textbf{chess}}
& SPMF MaxFEM & 17.19 / 1.00 & 459.9 / 1.00 & 164.70 / 1.00 & 487.1 / 1.00 & 427.05 / 1.00 & 472.9 / 1.00 \\
& Desbordante MaxFEM & 5.00 / 3.58 & \textbf{6.8 / 67.92} & 38.37 / 4.14 & \textbf{8.7 / 57.60} & 117.15 / 3.78 & \textbf{8.0 / 61.35} \\
& \textbf{Desbordante ParMaxFEM} & \textbf{1.33 / 15.01} & 43.1 / 18.37 & \textbf{22.56 / 10.58} & 165.3 / 5.03 & \textbf{39.20 / 11.87} & 553.2 / 1.08 \\
\cmidrule{1-8}

\multirow{3}{*}{\textbf{kosarak}}
& SPMF MaxFEM & 12.06 / 1.00 & 1686.7 / 1.00 & 142.53 / 1.00 & 1976.2 / 1.00 & 401.26 / 1.00 & 1998.8 / 1.00 \\
& Desbordante MaxFEM & 6.10 / 1.73 & \textbf{205.9 / 8.05} & 68.77 / 2.07 & \textbf{224.2 / 8.81} & 188.04 / 2.15 & \textbf{232.2 / 8.55} \\
& \textbf{Desbordante ParMaxFEM} & \textbf{2.44 / 3.81} & 434.7 / 3.83 & \textbf{15.03 / 9.51} & 581.7 / 3.45 & \textbf{30.79 / 13.10} & 1006.7 / 2.52 \\
\cmidrule{1-8}

\multirow{3}{*}{\textbf{mushrooms}}
& SPMF MaxFEM & 7.90 / 1.00 & 470.6 / 1.00 & 115.28 / 1.00 & 477.7 / 1.00 & 559.87 / 1.00 & 377.8 / 1.00 \\
& Desbordante MaxFEM & 2.09 / 4.28 & \textbf{8.4 / 56.36} & 31.81 / 3.66 & \textbf{8.6 / 55.84} & 139.03 / 4.03 & \textbf{9.3 / 40.76} \\
& \textbf{Desbordante ParMaxFEM} & \textbf{0.31 / 25.23} & 29.8 / 18.47 & \textbf{5.66 / 21.68} & 136.5 / 4.84 & \textbf{23.51 / 23.81} & 400.0 / 0.94 \\
\cmidrule{1-8}

\multirow{3}{*}{\textbf{OnlineRet.}}
& SPMF MaxFEM & 12.88 / 1.00 & 779.8 / 1.00 & 104.94 / 1.00 & 1086.5 / 1.00 & 418.37 / 1.00 & 1626.7 / 1.00 \\
& Desbordante MaxFEM & 5.17 / 2.48 & \textbf{72.6 / 10.77} & 41.44 / 2.54 & \textbf{93.1 / 11.85} & 172.93 / 2.45 & \textbf{117.9 / 14.36} \\
& \textbf{Desbordante ParMaxFEM} & \textbf{1.82 / 6.00} & 89.4 / 8.72 & \textbf{11.23 / 9.35} & 115.4 / 9.45 & \textbf{29.79 / 14.24} & 185.8 / 8.77 \\
\cmidrule{1-8}

\multirow{3}{*}{\textbf{PAMAP}}
& SPMF MaxFEM & 12.53 / 1.00 & 1948.6 / 1.00 & 163.45 / 1.00 & 4536.1 / 1.00 & 517.01 / 1.00 & 5360.5 / 1.00 \\
& Desbordante MaxFEM & 3.93 / 2.40 & \textbf{421.2 / 4.31} & 40.52 / 4.05 & \textbf{563.1 / 8.05} & 117.71 / 4.54 & \textbf{580.1 / 9.24} \\
& \textbf{Desbordante ParMaxFEM} & \textbf{2.49 / 3.97} & 652.2 / 2.82 & \textbf{8.72 / 18.44} & 1432.7 / 3.25 & \textbf{21.98 / 23.49} & 2907.0 / 1.85 \\
\cmidrule{1-8}

\multirow{3}{*}{\textbf{RecordLnk}}
& SPMF MaxFEM & 9.16 / 1.00 & 3216.4 / 1.00 & 149.36 / 1.00 & 3129.9 / 1.00 & 484.40 / 1.00 & 3797.2 / 1.00 \\
& Desbordante MaxFEM & 1.99 / 4.77 & \textbf{191.6 / 17.47} & 28.05 / 5.46 & \textbf{204.9 / 15.55} & 105.24 / 4.75 & \textbf{208.8 / 18.20} \\
& \textbf{Desbordante ParMaxFEM} & \textbf{1.06 / 8.58} & 687.2 / 5.15 & \textbf{6.93 / 21.82} & 967.3 / 3.72 & \textbf{20.87 / 23.76} & 933.1 / 4.08 \\
\cmidrule{1-8}

\multirow{3}{*}{\textbf{retail}}
& SPMF MaxFEM & 15.42 / 1.00 & 408.5 / 1.00 & 173.47 / 1.00 & 392.3 / 1.00 & 382.72 / 1.00 & 431.6 / 1.00 \\
& Desbordante MaxFEM & 6.80 / 2.31 & \textbf{25.3 / 16.12} & 71.10 / 2.48 & \textbf{32.6 / 12.15} & 148.25 / 2.63 & \textbf{41.6 / 10.37} \\
& \textbf{Desbordante ParMaxFEM} & \textbf{1.51 / 9.83} & 57.3 / 7.45 & \textbf{16.05 / 10.58} & 77.9 / 5.03 & \textbf{43.83 / 9.43} & 111.9 / 4.02 \\
\cmidrule{1-8}

\multirow{3}{*}{\textbf{Skin}}
& SPMF MaxFEM & 11.74 / 1.00 & 981.1 / 1.00 & 257.86 / 1.00 & 1354.6 / 1.00 & 490.98 / 1.00 & 1088.1 / 1.00 \\
& Desbordante MaxFEM & 3.10 / 3.83 & \textbf{53.4 / 18.70} & 59.75 / 4.38 & \textbf{64.3 / 21.79} & 130.41 / 3.76 & \textbf{71.2 / 15.28} \\
& \textbf{Desbordante ParMaxFEM} & \textbf{0.61 / 15.35} & 138.5 / 7.41 & \textbf{10.00 / 25.91} & 509.4 / 3.17 & \textbf{20.46 / 24.00} & 730.7 / 1.49 \\
\hline

\rowcolor{gray!15}
\cellcolor{gray!15}\multirow{3}{*}{\textbf{AVG / GEOMEAN}}
& \cellcolor{gray!15}SPMF MaxFEM
  & \cellcolor{blue!15} 13.70 / 1.00 & \cellcolor{red!15} 895.1 / 1.00
  & \cellcolor{blue!15} 144.71 / 1.00 & \cellcolor{red!15} 1240.5 / 1.00
  & \cellcolor{blue!15} 434.05 / 1.00 & \cellcolor{red!15} 1527.7 / 1.00 \\
\rowcolor{gray!15}
& \cellcolor{gray!15}Desbordante MaxFEM
  & \cellcolor{blue!15} 6.00 / 2.25 & \cellcolor{red!15} \textbf{102.8 / 13.61}
  & \cellcolor{blue!15} 52.60 / 2.88 & \cellcolor{red!15} \textbf{119.9 / 13.86}
  & \cellcolor{blue!15} 168.66 / 2.68 & \cellcolor{red!15} \textbf{176.1 / 11.43} \\
\rowcolor{gray!15}
& \cellcolor{gray!15}\textbf{Desbordante ParMaxFEM}
  & \cellcolor{blue!15} \textbf{2.15 / 5.89} & \cellcolor{red!15} 182.2 / 8.01
  & \cellcolor{blue!15} \textbf{13.95 / 11.77} & \cellcolor{red!15} 388.6 / 4.14
  & \cellcolor{blue!15} \textbf{41.44 / 11.86} & \cellcolor{red!15} 712.7 / 2.79 \\
\bottomrule

\multicolumn{8}{l}{\rule{0pt}{3ex}\textbf{Format:} \textit{Average Value} / \textit{Factor vs. SPMF} (values $>1$ indicate improvement).} \\
\multicolumn{8}{l}{\textbf{Metrics:} Time in seconds (Speedup); Memory in MB (Reduction Factor). \textbf{Bold} indicates the best result.} \\
\multicolumn{8}{l}{\textbf{Configuration:} Data points correspond to experiments with $winlen \in \{5, 10, 15\}$ and varying $minsup$ values selected to fit these specific time intervals.}
\end{tabular}
\end{table*}

\subsection{Experiments}

We utilized a diverse set of real-world datasets from the SPMF repository\footnote{\url{https://www.philippe-fournier-viger.com/spmf/index.php?link=datasets.php}}, covering various domains such as retail, web click-streams, and sensor data. The characteristics of these datasets are summarized in Table~\ref{tab:datasets}.

To answer the posed research questions, we ran the following experiments.

\textbf{Experiment 1.}
In this initial experiment, we replicated the evaluation scenario performed by the MaxFEM authors on the \textit{kosarak} and \textit{retail} datasets to benchmark our proposed Desbordante MaxFEM and Desbordante ParMaxFEM implementations against the original SPMF baseline. While we maintained the original window lengths ($winlen \in \{5, 10, 15\}$), we increased the granularity of our analysis by testing lower minimum support thresholds. This allowed us to assess the performance of all three implementations under heavier computational loads where execution times are substantially longer. Moreover, we provide a comparative analysis of memory consumption alongside runtime metrics. Finally, similar to the original experiment, we plot the number of discovered maximal episodes to contextualize the computational complexity relative to the output size.

\textbf{Experiment 2.}
To obtain a comprehensive performance landscape, we benchmarked SPMF MaxFEM, Desbordante MaxFEM, and Desbordante ParMaxFEM across all the aforementioned datasets. For each dataset, we varied the $minsup$ parameter while using the same set of window lengths as in Experiment 1. The $minsup$ values were carefully selected based on the baseline SPMF runtime to cover a wide range of execution scenarios, from approximately two seconds to ten minutes. ParMaxFEM was configured to utilize 8 threads. We report both absolute values (seconds, megabytes) and relative metrics (speedup factor for time, reduction factor for memory) compared to the SPMF baseline.

\textbf{Experiment 3.} This experiment studied the maximum achieved speed up by ParMaxFEM and Desbordante MaxFEM compared to SPMF MaxFEM. The data was taken from the same set of runs that was produced for Experiment 2.

\textbf{Experiment 4.} Finally, we studied ParMaxFEM scalability in our last experiment. We varied the number of threads in our thread pool and measured run time alongside with memory consumption on data from Experiment 2.

\subsection{Discussion}

\textbf{Experiment 1.} Fig. ~\ref{fig:kosarak-retail} shows run times, memory consumption and the number of found episodes for kosarak and retail datasets.

In this experiment we used plots to examine the detailed behavior across many data points. The plots show that Desbordante’s MaxFEM and ParMaxFEM are faster than SPMF’s MaxFEM and use less memory. They also show that increasing the $minsup$ and decreasing $winlen$ reduces both memory usage and runtime.

It is worth noting the occasional fluctuations in the SPMF memory measurements. These are artifacts of the JVM's memory management heuristics, which often pre-allocate heap space aggressively. Since our methodology utilizes \texttt{GNU Time} to capture the peak Resident Set Size (RSS), these values accurately reflect the total physical memory consumed by the process. We consider this the most practically relevant metric, as it represents the actual resource footprint imposed on the user's system, rather than a size of internal data structures.

We observe a clear linear relationship between execution time and the number of discovered maximal episodes with relatively low dispersion. The slope of this linear trend depends on the window size ($winlen$), indicating a relationship between $winlen$ and the linearity coefficient.

Thus, Desbordante's MaxFEM and ParMaxFEM extend the practical applicability of maximal frequent episode mining across a substantial portion of the minimal‑support and window‑size ranges, improving the method’s real‑world usefulness.

\begin{figure*}
    \centering
    \includegraphics[width=1\linewidth]{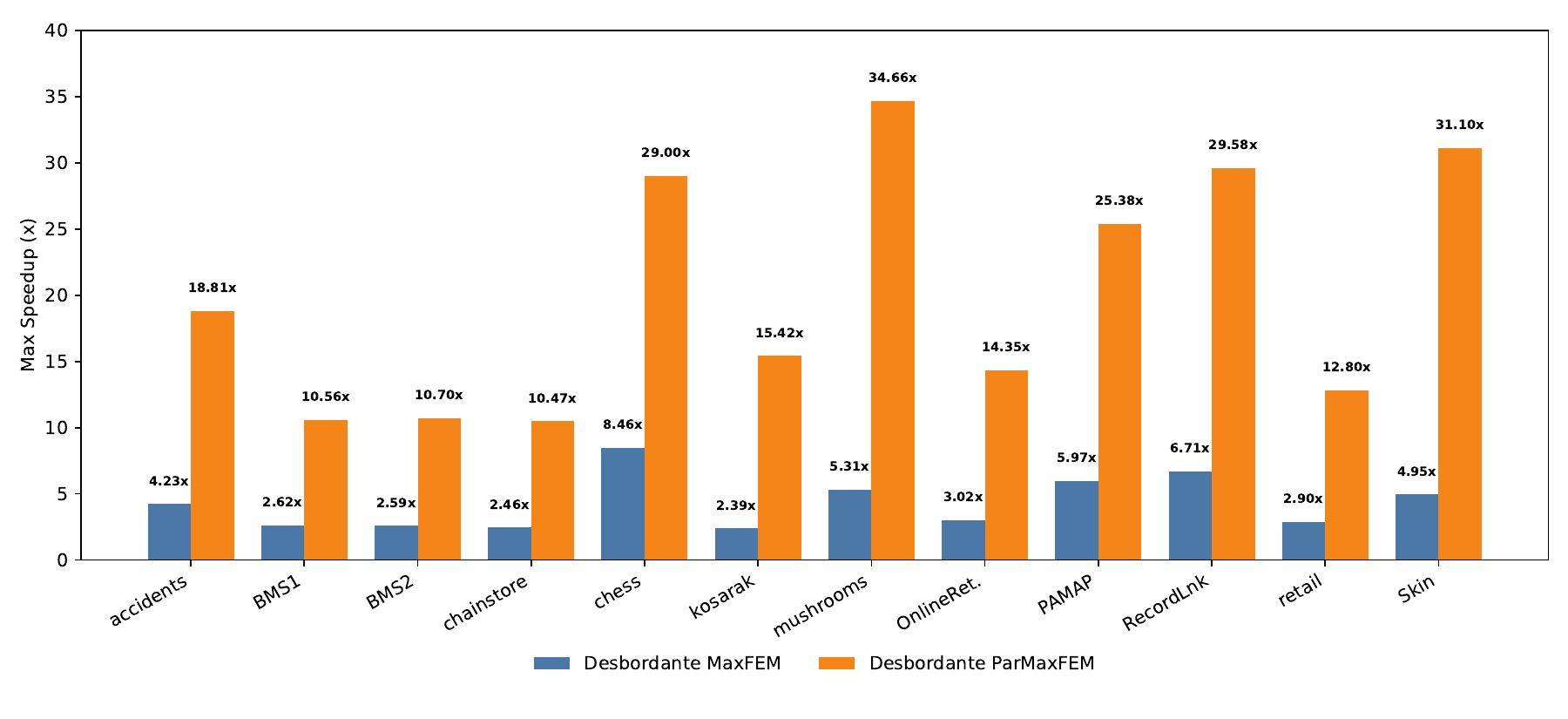}
    \caption{Maximum observed speedup achieved by Desbordante MaxFEM and Desbordante ParMaxFEM across all evaluated parameters for each dataset.}
    \label{fig:maximum}
\end{figure*}

\textbf{Experiment 2.}

Table~\ref{tab:perf_final_fixed_data} presents the overall performance landscape. Given the large number of data points resulting from various parameter combinations, we aggregated the results based on the execution time of the SPMF baseline. We classified the experiments into three workload groups: Small (SPMF runtime in $[2\text{s}, 60\text{s})$), Medium ($[1\text{m}, 5\text{m})$), and Large ($[5\text{m}, 10\text{m})$). For each dataset within a group, we report the arithmetic mean of the absolute values (execution time and peak memory). Additionally, we provide the geometric mean of the relative metrics (speedup and memory reduction factor), calculated point-by-point against the SPMF baseline.

It is important to note that identifying specific $minsup$ thresholds that yield execution times within these target intervals is a non-trivial task. The sensitivity of runtime to the $minsup$ parameter varies drastically across datasets. Sparse datasets, such as \textit{kosarak} and \textit{retail} (analyzed in Experiment 1), exhibit a relatively smooth and gradual growth in execution time as $minsup$ decreases. However, for many other datasets, particularly dense ones, the growth is much more abrupt.

For instance, in the \textit{chess} dataset (density 49.33\%, average transaction length 37), the computational complexity behaves explosively: a decrease in $minsup$ by a single unit (representing approx. 0.03\% of the dataset size) can result in a multi-fold increase in runtime. To efficiently locate the necessary data points under these conditions, we developed an adaptive search strategy. We started with a high $minsup$ and applied a geometric step-down (multiplying by a factor of 0.9). Upon triggering the time limit (TL), the algorithm switched to a binary search between the last successful value and the timeout value. This procedure was executed to find $minsup$ point per interval for each window length $winlen \in \{5, 10, 15\}$.

At the bottom of the table we present values averaged over all 12 datasets. The best value for a dataset is highlighted in bold.

As you can see, Desbordante MaxFEM is $1.40 - 4.54 \times$ faster than SPMF MaxFEM, depending on the dataset. Averaged over datasets the improvement ranged from $2.25$ to $2.68 \times$.

Memory consumption was impacted more profoundly. The reduction ranged from $4.31$ to $67.92 \times$, on average $11.43 - 13.61 \times$.

The results indicate that the careful reimplementation yielded clear benefits, especially in terms of memory consumption.

ParMaxFEM exhibits different behavior on our benchmark. First of all, its run times are drastically better and improving with moving to larger groups. The improvement ranged $2.44 - 25.23 \times$ depending on the dataset and averaged to $5.89 - 11.86 \times$ range over all datasets. However, memory consumption increased, compared to Desbordante MaxFEM, but it is still lower than SPMF MaxFEM. In this case the reduction ranged from $0.94$ to $18.37 \times$ and averaged to $2.79 - 8.01 \times$ over all datasets. It is interesting that memory savings decrease while moving to larger groups.

The overhead arises because each thread maintains its own data (composite episodes) in thread-local storage. During the parallelized step 5, multiple copies of identical episodes are created and must be tracked. Thus, the algorithm trades increased memory usage for higher speed. We study this behavior further in experiment 4.

\textbf{Experiment 3.} Fig. \ref{fig:maximum} illustrates the maximum observed speedup achieved by both the optimized sequential MaxFEM and the parallel ParMaxFEM (8 threads) across all 12 datasets, using the original SPMF implementation as the baseline.

The results show that performance gains are highly dependent on the dataset's characteristics. The highest speedups are observed in dense datasets with complex search spaces, such as \textit{mushrooms} (34.66$\times$), \textit{Skin} (31.10$\times$), and \textit{chess} (29.00$\times$).

Overall, our evaluation shows that ParMaxFEM provides a significant performance leap across the entire benchmark suite. The sequential C++ implementation provides a speedup ranging from 2.39$\times$ to 8.46$\times$, while the parallel version operating on 8 threads achieves a total speedup range of 10.47$\times$ to 34.66$\times$.

\textbf{Experiment 4.} Fig.~5 shows the average parallel scalability (geometric mean) of the algorithm on data from Experiment 2. As the number of threads increases, the average speedup falls below the ideal linear line ($y = x$), reaching approximately $8.8\times$ at 32 threads. This sublinear growth follows Amdahl's Law: fitting our data yields a parallelizable fraction of $P \approx 0.916$, setting a theoretical maximum speedup of roughly $11.9\times$. However, it is crucial to note that this overall trend hides significant variance: while some individual runs scale exceptionally well, others plateau much earlier.

Memory usage grows linearly with the number of threads. On average, it increases by a factor of $3.2\times$ at 32 threads, remaining below the SPMF baseline for most runs. However, this is an average estimate, as some datasets (e.g., \textit{mushrooms}) already reach this limit earlier.

\begin{figure}
    \centering
    \includegraphics[width=1\linewidth]{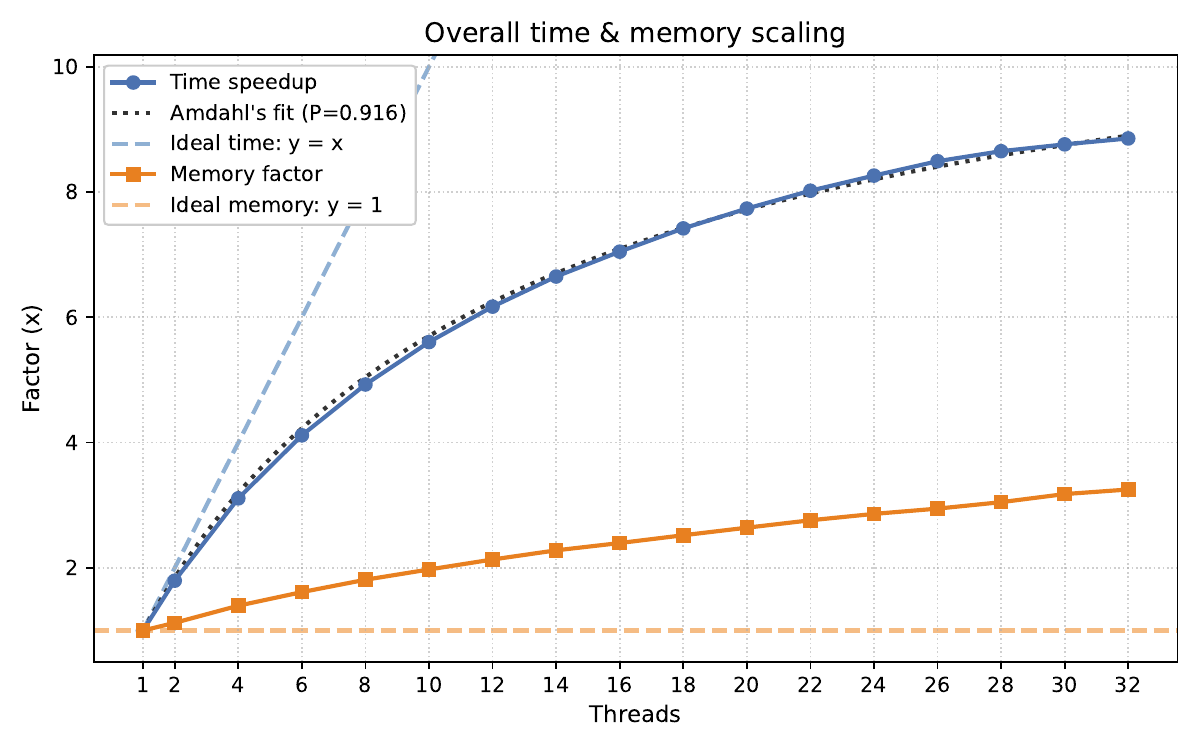}
    \caption{Performance and memory consumption scalability}
    \label{fig:scalability}
\end{figure}

\subsection{Wrap-up and Key Takeaways}

First of all, we must note that benchmarking these algorithms is challenging because performance depends heavily not only on the dataset characteristics, but also on the user-specified parameters: the window length and minimum support threshold. We selected datasets with diverse characteristics and varied parameters to cover a wide range of scenarios, which requires many experiments across many points. Now, turning to the posed research questions.

\textbf{Implementation Efficiency (RQ1).} Reimplementing in C++ yields a speedup of $1.40 - 4.54 \times$ due to system‑level optimizations, consistent with our prior work.

\textbf{Performance Superiority (RQ2)}. Parallelization provides substantial gains. ParMaxFEM is faster by $2.39 - 34.66\times$ compared to the reimplemented MaxFEM.

\textbf{Parallel Scalability (RQ3).} The algorithm on balanced data exhibits average scalability governed by Amdahl's Law ($P \approx 0.916$), implying a theoretical maximum speedup of roughly $11.9\times$, while achieving an average speedup of $8.8\times$ at 32 threads.

\section{Conclusion}\label{sec:conclusion}

In this paper, we presented ParMaxFEM, a high-performance parallel algorithm for mining maximal frequent episodes in complex event sequences. Our work addresses three main limitations of the existing MaxFEM implementation: slow execution speed, high memory usage, and lack of proper Python integration.

We tested our implementation on 12 real-world datasets from different domains. The experiments confirmed that our approach works well across diverse data types and parameter ranges. On 8 CPU cores, ParMaxFEM achieves an average speedup of 11$\times$ compared to the original MaxFEM implementation, with maximum speedup reaching up to 35$\times$ on some datasets. Our work makes maximal frequent episode mining more practical for real-world applications. The improved performance allows users to analyze larger datasets and explore more parameter combinations in reasonable time, including mining episodes with lower support thresholds and longer window sizes that were previously too computationally expensive.

We added it to Desbordante, an \textit{open-source} \textit{high-performance} data profiler, while providing a native Python interface that allows users to work with event sequences directly from Python without file conversion or subprocess overhead. The ParMaxFEM implementation is available as a pull request\footnote{\url{https://github.com/Desbordante/desbordante-core/pull/676}}, that is currently under review and will be merged into the main Desbordante repository. The Python package will be available through PyPI, making it easy for data scientists to install and use ParMaxFEM in their projects.

With the introduction of ParMaxFEM, Desbordante begins to support sequences as a core data type. We plan to extend this further~--- additional sequence-based tasks and algorithms will be added in the future.

\bibliographystyle{IEEEtran}

\bibliography{my}

\end{document}